\documentclass[letterformat]{emulateapj}

\shorttitle{C\,{\sc ii} abundances in early-type stars}
\shortauthors{Nieva \& Przybilla}

\begin{document}

\title{C\,{\sc ii} abundances in early-type stars: Solution to
a notorious non-LTE problem}

\author{M. F. Nieva}
\affil{Dr. Remeis Sternwarte Bamberg, Sternwartstr. 7, D-96049 Bamberg, Germany}
\affil{Observat\'orio Nacional, Rua General Jos\'e Cristino 77 CEP 20921-400, Rio de Janeiro, Brazil}
\email{nieva@sternwarte.uni-erlangen.de}

\and

\author{N. Przybilla}
\affil{Dr. Remeis Sternwarte Bamberg, Sternwartstr. 7, D-96049 Bamberg, Germany}
\email{przybilla@sternwarte.uni-erlangen.de}

\begin{abstract}
We address a long-standing discrepancy between non-LTE analyses of
the prominent C\,{\sc ii} $\lambda\lambda$4267 and 6578/82\,\AA~multiplets
in early-type stars. A comprehensive non-LTE model atom of C\,{\sc ii} is
constructed based on critically selected atomic data. This model atom is
used for an abundance study of six apparently slow-rotating main-sequence
and giant early B-type stars. High-resolution and high-S/N spectra
allow us to derive highly consistent abundances 
not only from the classical features but also from up to 18 further C\,{\sc ii}
lines in the visual -- including two so far unreported emission
features equally well reproduced in non-LTE.
These results require the stellar atmospheric parameters to be determined
with care.
A homogeneous (slightly) sub-solar present-day carbon abundance from young stars in the solar
vicinity (in associations and in the field) of 
$\log$\,C/H\,$+12=8.29\pm0.03$ is indicated.
\end{abstract}



\section {Introduction}

One of the most abundant metals in the universe is carbon,
the central building block of all organic chemistry.
Abundances derived from luminous early B- and late O-type stars provide important 
constraints on stellar and galactochemical evolution. 
In extragalactic applications (e.g. analyses of dwarf stars in the Magellanic
Clouds, see e.g. Hunter et al.~2005;
Korn et al.~2005) one desires to 
study the {\em strongest} features in the metal line spectra, as these are the
only measurable at low S/N and/or high projected rotational
velocities. In the case of ionized carbon these are the prominent lines of the two 
multiplets C\,{\sc ii} $\lambda\lambda$4267.02/4267.27\,\AA~and
6578.03/6582.85\,{\AA}. These lines are unfortunately highly sensitive to
non-LTE effects, as well as to the choice of stellar atmospheric parameters.
So far, all studies from the literature failed to derive consistent abundances 
from these lines.

The problem was addressed by Lennon~(1983), Eber \& Butler~(1988) and
Sigut~(1996:~S96) using C\,{\sc ii} model atoms of increasing complexity.
Spectrum synthesis based on the latter two non-LTE model atoms obtained better -- 
but apparently still not good -- agreement with observation when compared to 
earlier work or LTE results. Non-LTE abundance analyses in Galactic 
OB stars e.g.~by Gies \& Lambert~(1992) and\,Kilian~(1992)
employing the Eber \& Butler (1988) model derived a metal deficiency (including
C) in these young stars with respect to the Sun, in accordance with
studies of H\,{\sc ii}~regions.

The present work aims at providing a solution to the classical non-LTE problem of
carbon abundance determinations in OB stars. A reliable C\,{\sc ii} model atom
is developed and first applications on high-quality spectra are presented. 
Besides great care in the selection of atomic data, special emphasis is also
given to an accurate atmospheric parameter determination, both in order to 
minimize systematic uncertainties. The measurable C\,{\sc ii} spectrum in
the visual is investigated, the two prominent line multiplets as well as
numerous weaker lines.

\section {Model calculations}

A hybrid approach is used for the non-LTE line formation computations. These are
based on line-blanketed 
LTE model atmospheres calculated with ATLAS9 (Kurucz~1993). Synthetic
spectra are computed with recent versions of DETAIL and SURFACE
(Giddings~1981; Butler \& Giddings~1985), solving the 
restricted non-LTE problem -- see Przybilla et al.~(2001) for details.

Hydrogen and He\,{\sc i/ii} populations are computed using recent model
atoms by Przybilla \& Butler~(2004) and Przybilla~(2005), respectively.
We have compared our synthetic H and He\,{\sc i/ii} lines 
with predictions from Lanz \& Hubeny~(2003) 
in the effective temperature range of 
27500\,K\,$\leq$\,$T_\mathrm{eff}$\,$\leq$\,32500\,K 
for dwarf as well as giant stars. Overall good agreement is
found. Exceptions are the He\,{\sc i} singlet lines, which are predicted
(significantly) stronger in our approach, in accordance with observation.

The C\,{\sc ii} model atom considers LS-terms up to principal quantum number
$n$\,$=$\,10 and angular momentum $\ell$\,$=$\,9
explicitly in non-LTE, with all
fine-structure sub-levels combined into one. 
The doublet~and
quartet spin-systems are treated simultaneously.
Level energies are adopted from Moore~(1993), S96 and Quinet~(1998).

Oscillator strengths from three sources are considered:
fine-structure data from {\em ab-initio} computations using the multiconfiguration
Hartree-Fock method in the Breit-Pauli approximation (Froese Fischer \&
Tachiev~2004: FFT04), data from an application of the Breit-Pauli
R-matrix method (Nahar~2002a) and results obtained in the
Opacity Project (OP) from a close-coupling method in the LS-approximation
(Yan, Taylor \& Seaton~1987). Our primary source of $gf$-values is FFT04,
followed by OP and Nahar for the remaining
transitions. Note that data for several important transitions from Nahar
disagree with those from FFT04 and OP, which show consistency among each other.
Intercombinations are neglected because of the very small
$gf$-values and high densties.

Photoionization cross-sections are taken from the OP (Yan \& Seaton~1987) 
where available.
For the remain\-der, data from Nahar~(2002b) are adopted. 
The choice is empirically motivated, giving preference to the OP data in order
to reproduce observation over the entire parameter range simultaneously
from all indicators. The
two data sets show differences in the resonance structures.

Effective collision strengths for electron impact excitation among
the lowest 16 LS-states are adopted from R-matrix computations of Wilson,
Bell \& Hudson~(2005).
Collisional excitation for transitions without reliable data are treated using
the Van Regemorter~(1962) approximation in the optically
allowed case and via the semi-empirical Allen~(1973) formula otherwise. 
Collision strengths varying between 0.01 ($\Delta
n\,\geq\,4$) to 100 ($\Delta n\,=\,0$) are employed, as suggested by evaluation of
the detailed data 
of Wilson et al.~(2005).

Collisional ionization rates are evaluated according to the Seaton~(1962) 
approximation, with threshold photoionization cross-sections
from OP and Nahar~(2002b), allowing for an empirical correction of one order of
magnitude higher for the 6f\,$^2$F$^{\circ}$ 
and 6g\,$^2$G terms -- corresponding to the upper levels of the
C\,{\sc ii} $\lambda\lambda$6151 and 6462\,\AA~transitions. 
For completeness, a C\,{\sc iii} model atom is also accounted for in the
computations, but its details are of no further importance to the present
study.

Voigt profiles are adopted in the formal solution using SURFACE.
Wavelengths and $gf$-values are taken from Wiese et al.~(1996). 
Radiative damping parameters are calculated from OP lifetimes and coefficients for
collisional broadening by electron impact are adopted from Griem~(1974)
for the C\,{\sc ii} $\lambda$4267\,\AA~doublet, while the approximation of 
Cowley~(1971) is used for the other lines.

\section {Analysis}

Six apparently slow-rotating stars are considered for the model atom
calibration and first applications. The observations consist of 
high-resolution, high-S/N spectra with wide wavelength
coverage, obtained with FEROS at the 2.2m telescope at ESO (La Silla, Chile).

\begin{figure}
\centering
\includegraphics[width=.92\linewidth]{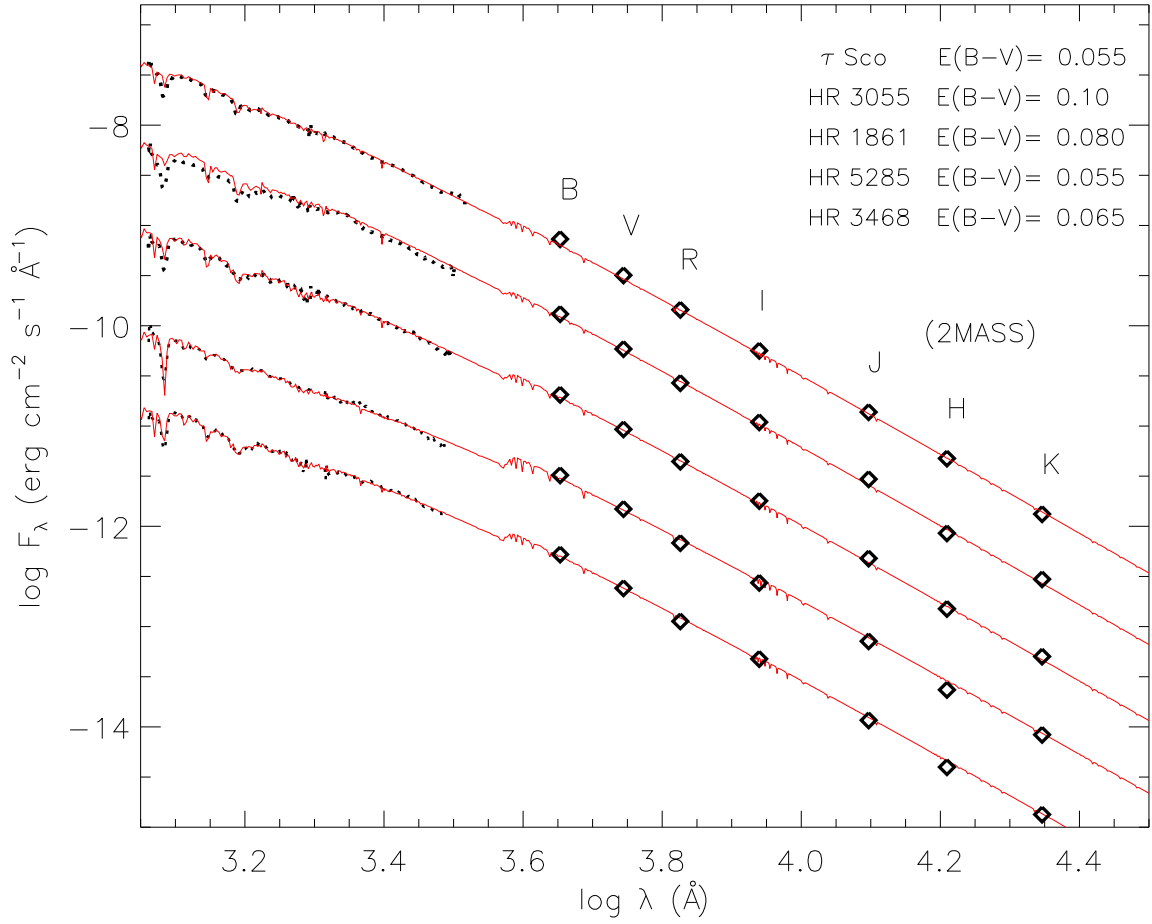}
\caption{Best fits of theoretical energy distributions to measurements by IUE
  (dotted lines)
  and Johnson and near-IR 2MASS photometry (diamonds). 
  The observed fluxes are dereddened by the values indicated using a 
  standard reddening law and 
  assuming $R{_V}=A{_V}/E(B-V)=3.1$ as typical for the local ISM. 
  They were degraded to the resolution of the ATLAS9 fluxes. 
  The models are normalized to the observed V magnitudes and
  shifted for clarity relative to each other. See Fig.~\ref{fig5} for
  atmospheric parameters.}
\label{fig1}
\end{figure}

\begin{figure}
\centering
\includegraphics[width=.86\linewidth]{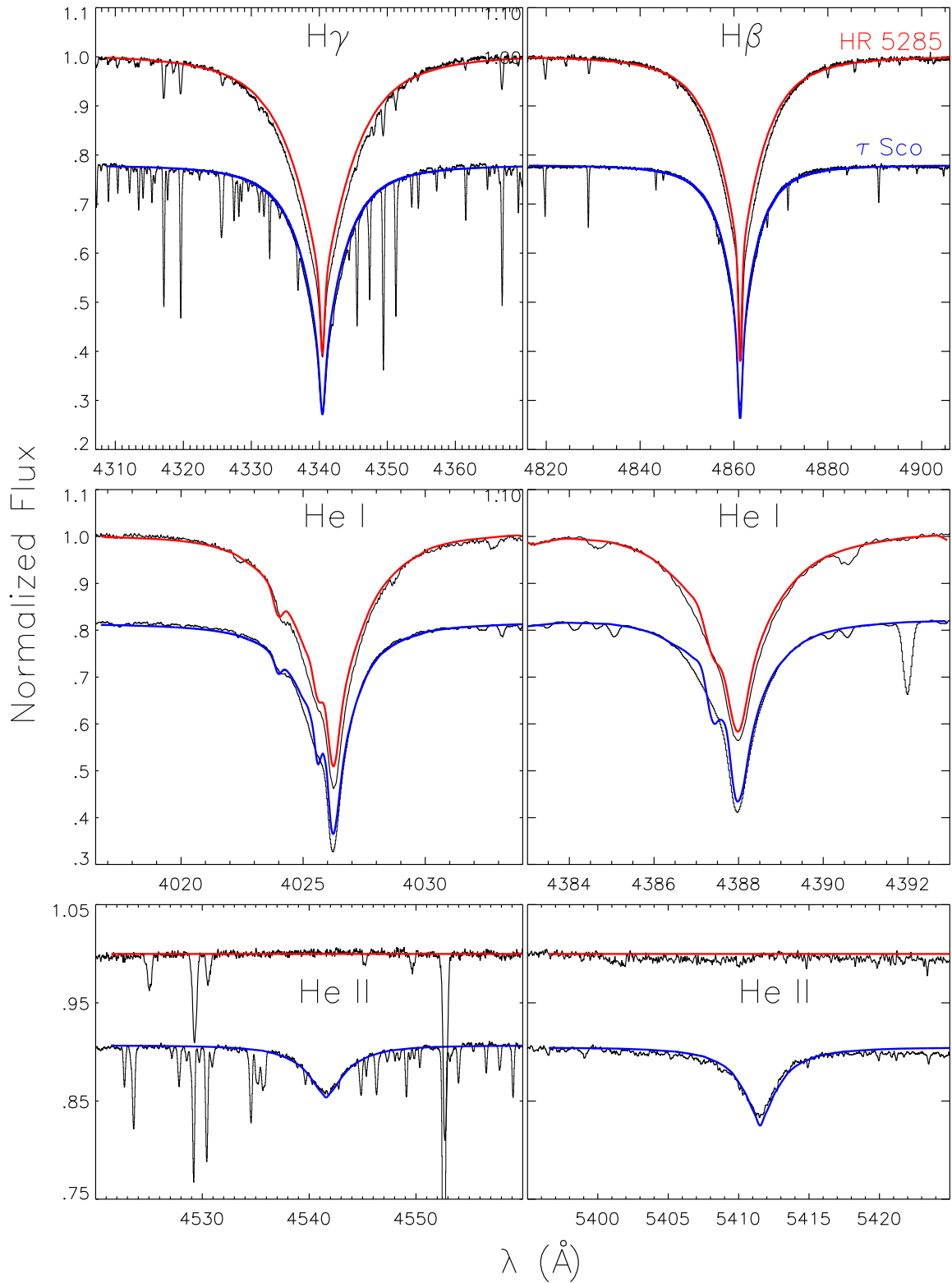}
\caption{Comparison of synthetic H and He\,{\sc i/ii} lines (smooth
  lines) with observation of a B0\,V ($\tau$\,Sco) and a B2\,V (HR\,5285) star. 
  Atmospheric parameters
  as summarized in Fig.~\ref{fig5} are 
  employed.}
\label{fig2}
\end{figure}

Effective temperatures $T_\mathrm{eff}$ are derived spectrophotometrically 
from IUE fluxes and Johnson and 2MASS magnitudes for all the stars except 
HR\,2928  where $T_\mathrm{eff}$ is in agreement with Kilian~(1992). Further
constraints can be derived from the He\,{\sc i/ii} ionization equilibrium for
$\tau$\,Sco and HR\,3055. Figure~\ref{fig1} displays best fits to
the spectrophotometry. Gravities $\log g$ are derived from line
profile fits to H$\delta$, H$\beta$, H$\gamma$ and H$\alpha$. 
Examples for some H and He\,{\sc i/ii} lines in $\tau$\,Sco and HR\,5285
are given in Fig.~\ref{fig2}.

After constraining $T_\mathrm{eff}$ and $\log g$ we compute small grids of
synthetic spectra for different carbon abundance
$\varepsilon$(C)=~$\log(N_{\rm C}/N_{\rm H})\,+\,12$ and 
several values of microturbulence $\xi$.
These grids are used for C abundance determinations from 
observation via a $\chi^{2}$-minimization technique. 
The free fitting parameters are $\varepsilon$(C), $v \sin i$ and
macroturbulence $\zeta$. The macroturbulent velocities in the sample stars
remain smaller than twice the sound speed.
The microturbulent velocity is fixed in
the usual approach, by demanding $\varepsilon$(C) to be independent of the
equivalent width W$_{\rm\lambda}$ of the C\,{\sc ii} line ensemble (see Fig.~\ref{fig5}).

\begin{figure}
\begin{center}
\includegraphics[width=\linewidth]{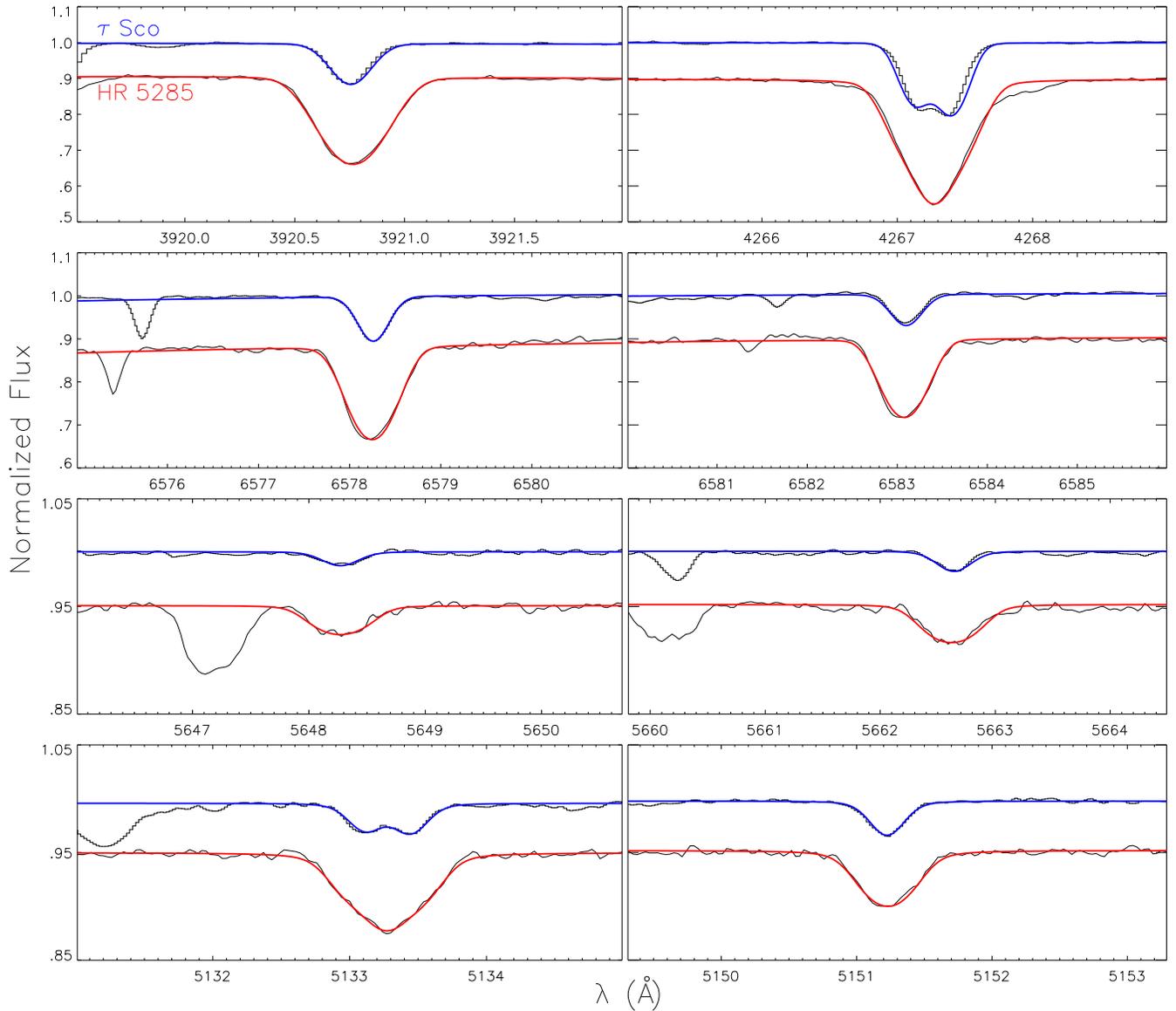}\\
\end{center}
\caption{Examples of best fits (smooth lines) to observed C\,{\sc ii} features for a 
  B0\,V ($\tau$\,Sco) and
  a B2\,V (HR\,5285) star. Atmospheric parameters and averaged $\varepsilon$(C) of
  each star are given in Fig.~\ref{fig5}.}
\label{fig3}
\end{figure}

\begin{figure}
\centering
\includegraphics[width=.85\linewidth]{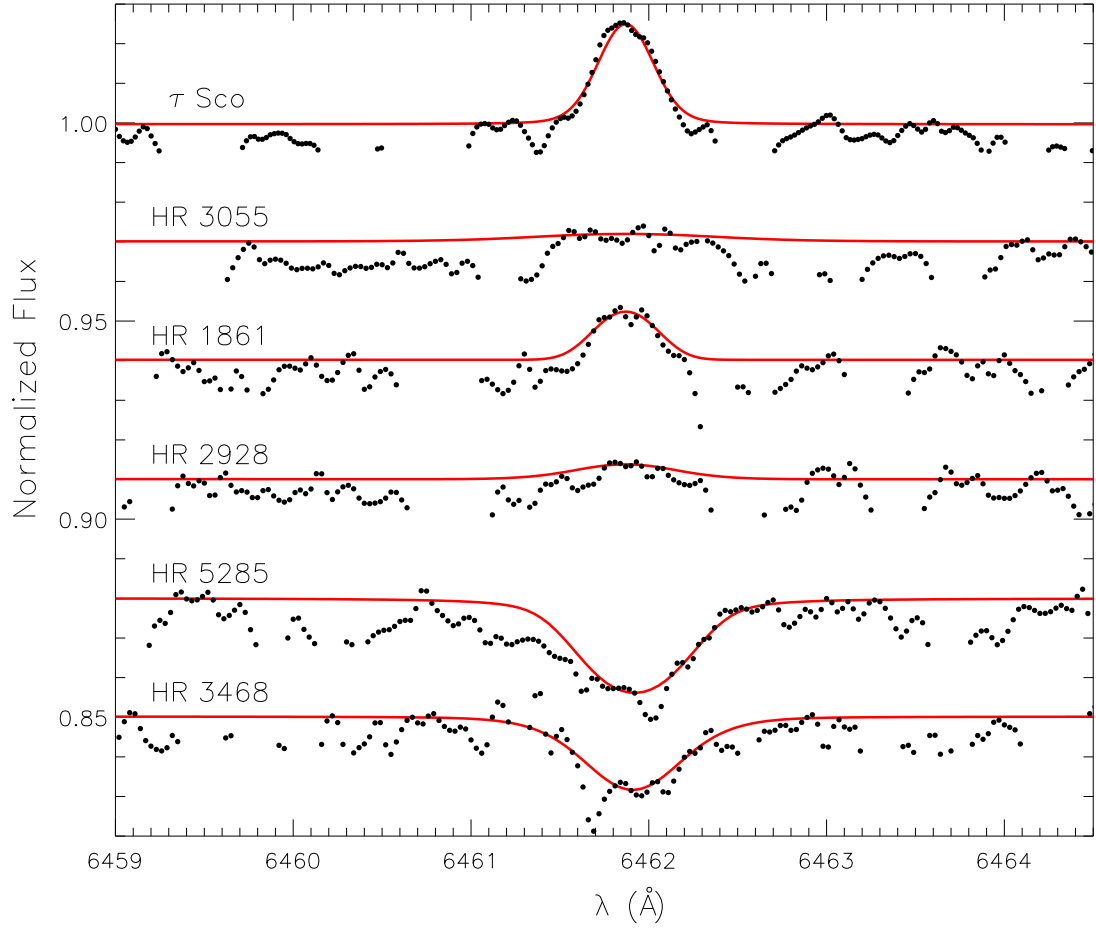}
\caption{Best fits to the C\,{\sc ii} $\lambda$6462\,\AA~line in the six
sample stars. For clarity, the stronger telluric features
in this spectral region are suppressed in the observed spectra (dots).}
\label{fig4}
\end{figure}

An excellent match between theory and observation is achieved, as shown in 
Fig.~\ref{fig3} exemplarily for the hottest ($\tau$\,Sco) and coolest
(HR\,5285) star in our sample. 
The complete linelist includes the C\,{\sc ii} $\lambda\lambda$3919.0/20.6,
4267.0/.2, 5133.3/37.3/39.2/43.4/45.2/51.1, 5648.1/62.5,
6151.5, 6461.9,
6578.0/82.9, 6779.9/80.6/83.1/87.2/6791.5 and 6800.7\,\AA~transitions.
Fits to C\,{\sc ii} $\lambda$6462\,\AA~ for all the stars are shown in
Fig.~\ref{fig4}. This line together with C\,{\sc ii} $\lambda$6151\,\AA~ is
subject to marked non-LTE effects, turning from absorption at spectral type
B2 into emission at earlier spectral types, a behaviour never reported
before. The emission results from a non-LTE overpopulation of the upper levels
of these transitions relative to the lower levels, facilitated by close
collisional coupling of the former to the C\,{\sc iii} ground state, which is close to LTE in all the sample stars.
For HR\,3055 and HR\,2928 C\,{\sc ii}~$\lambda\lambda$6462 and 6151\,\AA~are
not considered in deriving the average $\varepsilon$(C).
See S96 for a discussion of the nature of the non-LTE effects of C\,{\sc
ii}~$\lambda\lambda$4267 and 6578/82\,{\AA}.

Non-LTE and LTE abundances for all individual lines are displayed as a
function of
$W_{\lambda}$ in Fig.~\ref{fig5}, showing excellent consistency in non-LTE. A slight
degradation of the overall consistency is indicated for $\tau$\,Sco (see
below). Atmospheric
parameters and averaged $\varepsilon$(C) are also given.

\section{Discussion \& Results}
\begin{figure}
\centering
\includegraphics[width=\linewidth]{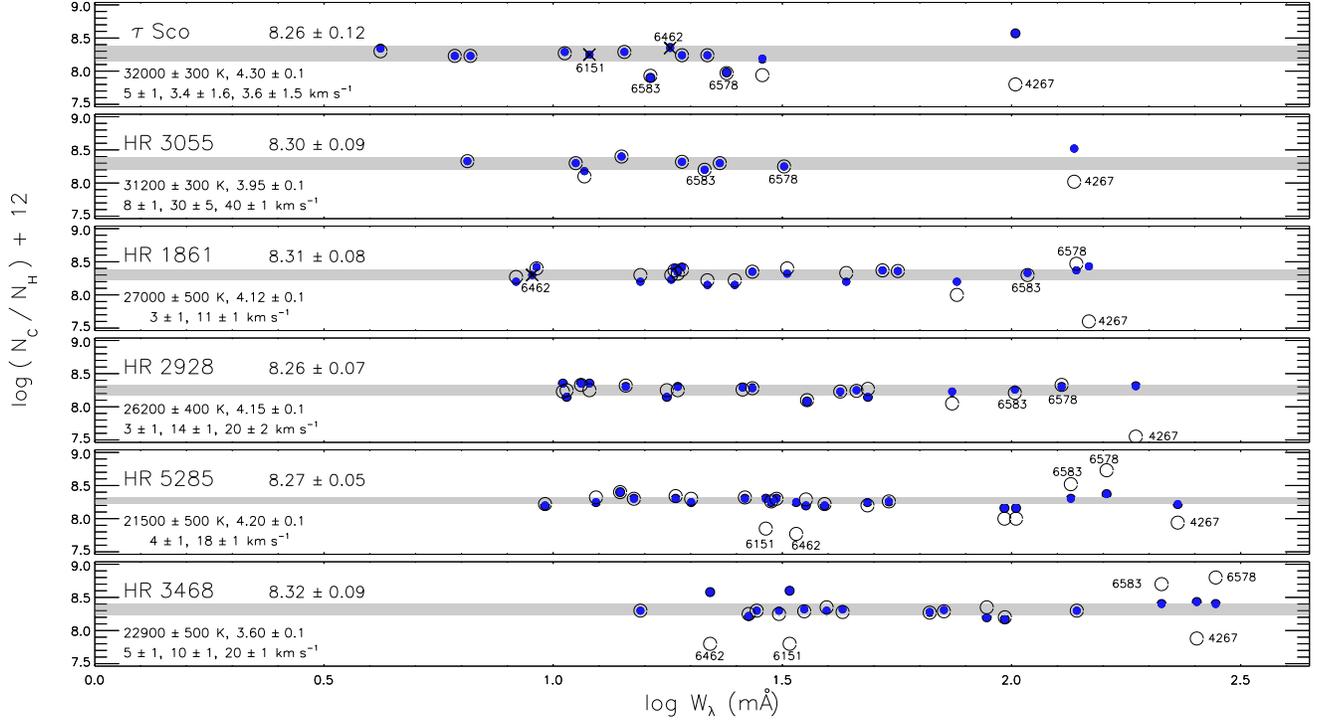}
\caption{Non-LTE (filled circles) and LTE (open circles) abundances
  vs. equivalent width for all the lines that could be measured in each
  spectrum (a linelist is provided in the text). In the upper left part of each
  row an identification of the star and its carbon abundance is given. 
  In the lower left part $T_\mathrm{eff}$,
  $\log g$, microturbulence, $v\sin i$ and macroturbulence (where non-zero)
  are shown. For abundances and velocities statistical 1$\sigma$-uncertainties
  are provided and for $T_\mathrm{eff}$ and $\log g$ the errors from our
  derivation are presented. The grey rectangles correspond to 1$\sigma$-uncertainties of
  $\varepsilon$(C). Identification of lines with high sensitivity to
  non-LTE effects is displayed. Emission lines are marked by crosses 
  (C\,{\sc ii} $\lambda\lambda$6151 and 6462\,\AA~in $\tau$\,Sco and
  $\lambda$6462\,\AA~in HR\,1861): LTE calculations are not able to reproduce them.}
\label{fig5}
\end{figure}

We have investigated the sensitivity of the transitions to modifications of atmospheric 
and several atomic parameters qualitatively. The individual lines
show a different behaviour: C\,{\sc ii} $\lambda\lambda$6151
and 6462\,\AA~are very sensitive to changes of $\log g$, $\xi$ and collisional
and photoionization cross-sections, 
C\,{\sc ii} $\lambda$4267\,\AA~is sensitive to photoionization, 
and all C\,{\sc ii} line strengths react sensitively on 
$T_\mathrm{eff}$ changes. 
As an example, exchanging our current~photo\-ionization data with the homogeneous
set of N02a would result in considerably reduced 
uncertainties in the mean $\varepsilon$(C) of $\tau$ Sco and HR 3055. 
However, for the same model configuration abundances from C\,{\sc ii} 
$\lambda$4267\,\AA~in the other 
stars are reduced by up to 0.5\,dex, approximately reproducing the LTE results in
Fig.~\ref{fig5}, while the remaining lines behave similarly to the final model.
Further analyses have to be made to quantify these~dependencies.

LTE analyses 
may produce abundances from the prominent C\,{\sc ii} $\lambda$4267\,\AA~transition in error 
by $\sim$0.3--0.8\,dex, by up to $\sim$0.4\,dex in the case of C\,{\sc ii}
$\lambda\lambda$6578/82\,\AA (note that $\sim$zero corrections may also
occur), and by $\sim$0.6--0.8\,dex for the weak C\,{\sc ii}
$\lambda\lambda$6151 and 6462\,\AA~lines in the cooler stars. In the
hotter stars the last two lines turn into emission and cannot be
reproduced at all assuming LTE. All other transitions are
subject to non-LTE corrections on the order of $\sim$0--0.2\,dex.

Noteworthy (small) discrepancies to the overall excellent agreement arise only for 
the C\,{\sc ii} $\lambda\lambda$4267 and 6578/82\,\AA~transitions in $\tau$\,Sco.
Among the sample stars $\tau$\,Sco is the only object showing 
a considerable (clumped) stellar wind and hard X-ray emission (see
Howk et al.~2000). The problems may be related to these complications,
as wind emission affects the H$\alpha$ profile and the
X-ray emission the photoionization rates, which both have to be modelled
correctly in order to reproduce these strongly non-LTE affected 
C\,{\sc ii} lines accurately.

Our non-LTE computations reproduce the 
C\,{\sc ii} $\lambda$4267\,\AA~theoretical equivalent widths of S96 (his Fig.~1). 
For C\,{\sc ii} $\lambda\lambda$6578/82\,\AA~we 
reproduce the S96 values at 15 kK; however at 20 kK our $W_{\lambda}$ are
$\sim$10\% lower and at 30 kK up to 50\% higher (see S96,
Figs.~5 and 6). The difference arises because we use non-LTE populations
when computing the H$\alpha$ line opacities which define the continuum
against which these lines are measured (S96 assumes LTE).
Note also that the $T_\mathrm{eff}$-scale 
employed by S96 for his comparison with observation appears to
produce values of $T_\mathrm{eff}$ {\em lower} by 700\,--\,3\,000\,K than our
derivations.

A highly consistent mean 
$\varepsilon$(C)\,$=$ 8.29$\pm$0.03 is derived from the sample stars,
which provides a tight estimate to the present-day C abundance from young
stars in the solar vicinity. The atmospheric composition appears to be unaffected 
by chemical mixing in the course of stellar evolution, i.e. we find no trend
of $\varepsilon$(C) with evolutionary age.~For comparison, adopting
Kilian's~(1992) results one derives a mean $\varepsilon$(C)\,$=$\,8.19$\pm$0.12 
from the same six stars, implying a systematic shift and
a significantly increased statistical scatter. More objects need to be further 
analyzed in order to verify the claim of such homogeneous present-day (slightly) sub-solar
-- considering as references $\varepsilon$(C)$_{\sun}$\,=\,8.39$\pm$0.05 
(Asplund et al.~2005) or $\varepsilon$(C)$_{\sun}$\,=\,8.52$\pm$0.06 
(Grevesse \& Sauval 1998) --
C abundances in nearby associations (HR\,1861: Ori OB1; $\tau$\,Sco,
HR\,5285: Sco-Cen) and in the field (the~other~stars).

\begin{acknowledgements}
We wish to thank K.~Butler for providing~DETAIL and SURFACE and M.~Altmann 
for kindly providing the FEROS data. We thank K.~Cunha and U.~Heber for their support. 
M.F.N acknowledges a scholarship by DAAD. 

\end{acknowledgements}

\clearpage

\clearpage

\clearpage

\clearpage

\clearpage

\end{document}